\documentclass[aps,twocolumn,prl,superscriptaddress,floatfix,amsmath,amssymb,showpacs,tightenlines]{revtex4}
\usepackage{epsfig,graphicx,times}
\usepackage{amstext}
\usepackage{amsmath}            
\usepackage{amssymb}            
\usepackage{graphicx}           
\usepackage{latexsym}
\usepackage{bm}

\begin{document}
\title{Controllable Coupling between Flux Qubits}

\author{ Yu-xi Liu}
\affiliation{Frontier Research System,  The Institute of Physical
and Chemical Research (RIKEN), Wako-shi 351-0198, Japan}
\author{L. F. Wei}
\affiliation{Frontier Research System,  The Institute of Physical
and Chemical Research (RIKEN), Wako-shi 351-0198, Japan}
\affiliation{IQOQI, Department of Physics, Shanghai Jiaotong
University, Shanghai 200030, P.R. China }
\author{J. S. Tsai}
\affiliation{Frontier Research System,  The Institute of Physical
and Chemical Research (RIKEN), Wako-shi 351-0198, Japan}
\affiliation{NEC Fundamental Research Laboratories, Tsukuba
Ibaraki 305-8501, Japan}
\author{Franco Nori}
\affiliation{Frontier Research System,  The Institute of Physical
and Chemical Research (RIKEN), Wako-shi 351-0198, Japan}
\affiliation{MCTP, Physics Department, CSCS, The University of
Michigan, Ann Arbor, Michigan 48109-1040, USA}
\date{\today}

\begin{abstract}
We propose an experimentally realizable method to control the
coupling between two flux qubits. In our proposal, the bias fluxes
are always fixed for these two inductively-coupled qubits. The
detuning of these two qubits can be initially chosen to be
sufficiently large, so that their initial interbit coupling is
almost negligible. When a time-dependent magnetic flux (TDMF) is
applied to one of the qubits, a well-chosen frequency of the TDMF
can be used to compensate the initial detuning and to couple two
qubits. This proposed method avoids fast changes of either qubit
frequencies  or the amplitudes of the bias magnetic fluxes through
the qubit loops, and also offers a remarkable way to implement any
logic gate as well as tomographically measure flux qubit states.

\end{abstract}

\maketitle \pagenumbering{arabic}

{\it Introduction.---} Superconducting Josephson junction circuits
currently provide one of the best qubit candidates, and
experiments have been performed for charge, flux, phase, and
charge-flux~\cite{physics} qubits. Quantum coherent oscillations
and conditional gate operations have been demonstrated using
two-coupled superconducting charge qubits~\cite{pashkin}. Further,
entangled macroscopic quantum states have been experimentally
verified in systems of coupled flux~\cite{izmalkov}, and
phase~\cite{xu,berkley} qubits.

Quantum computing requires that the interaction between different
qubits can be selectively switched on and off. This is an
extremely difficult and important issue. Several schemes have been
proposed to realize controllable couplings and local qubit
operations. One is a controllable coupling by dynamically tuning
the qubit frequencies, e.g., in
Refs.~\cite{berkley,majer,mcdermott,blais,cleland}. This tunable
proposal requires that different qubits have the same frequencies
(i.e., resonant interaction) when they are coupled. When they are
decoupled, one of their frequencies should be suddenly changed by
an external bias variable such that two coupled subsystems have a
larger detuning (i.e., non-resonant interaction). The second
approach uses switchable couplings in charge qubit circuits by
changing the bias magnetic flux, e.g., in
Refs.~\cite{makhlin,you}. In practice, the switching time of the
magnetic flux should be less than the inverse single-qubit
Josephson energy (less than a nanosecond), which is a challenge
for  present experiments. The third proposal requires additional
subcircuits, e.g., in Refs.~\cite{averin,plourde}. These
additional elements increase the complexity of the circuits and
add additional uncontrollable noise.

To easily switch on and off the coupling among qubits is one of
the most important open problems in quantum information hardware.
Here, we propose a way to overcome this severe problem plaguing
experiments. Specifically, we present a proposal on how to achieve
a controllable interaction between flux qubits by virtue of
time-dependent magnetic fluxes (TDMFs). Here, we make the same
assumption as in the decoupling
experiments~\cite{berkley,majer,mcdermott}, which require the two
qubits to be in the large detuning regime. However, here, the
two-qubit coupling and decoupling are controlled by {\it the
frequency} (not the dc component) of the applied TDMF. So we
completely avoid having to quickly change the bias magnetic
flux---a severe problem faced by many previous proposals for
superconducting qubits. Moreover, our proposal does not require
additional subcircuits. These merits also make our proposal
potentially useful for a variety of other types of qubit
experiments, and could solve a central problem in this field.

\begin{figure}
\includegraphics[bb=81 445 503 740, width=4.0 cm, clip]{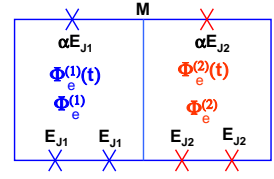}
\caption[]{(Color online) Two superconducting flux qubits are
coupled through their mutual inductance $M$. Each qubit loop
includes three junctions. The tunable interaction between two
qubits can be realized by changing the frequency of the external
magnetic flux $\Phi^{(l)}_{e}(t)\,(l=1,\,2)$ through the $l$th
qubit.}\label{fig1}
\end{figure}

{\it Controllable Hamiltonian.---} Two flux-quibts interact with
each other through a mutual inductance $M$, as shown in
Fig.~\ref{fig1}. Each qubit loop contains three junctions, one of
them has an area $\alpha$ times smaller than that of the two
identical junctions. The larger junction in the $l$th qubit loop
has Josephson energy $E_{{\rm J},l}$ ($l=1,\,2$). The
gauge-invariant phases (of the two identical junctions and the
smaller one) in the $l$th qubit are $\varphi^{(l)}_{1}$,
$\varphi^{(l)}_{2}$, and $\varphi^{(l)}_{3}$. We assume that a
static (dc) magnetic flux $\Phi^{(l)}_{\rm e}$ and a
time-dependent magnetic flux (TDMF) $\Phi^{(l)}_{\rm
e}(t)=A_{l}\cos (\omega^{(l)}_{c}t)$ are applied through the $l$th
qubit.  Using the phase constraint condition through the $l$th
qubit loop $\sum_{i=1}^{3}\varphi^{(l)}_{i}+(2\pi\Phi^{(l)}_{\rm
e}/\Phi_{0})+ (2\pi\Phi^{(l)}_{\rm e}(t)/\Phi_{0})=0$,  the total
Hamiltonian of the two qubits can be written as~\cite{neglect}
\begin{equation}\label{eq:1}
H=\sum_{l=1}^{2}(H_{l}+H^{(l)}_{\rm D})+\sum_{l\neq m=1}^2
H_{lm}+H_{C}+H_{A}.
\end{equation}
Here, the single qubit Hamiltonian is
$H_{l}=P^2_{P,l}/2M_{P,l}+P^2_{Q,l}/2M_{Q,l} +2E_{{\rm
J},l}(1-\cos\varphi^{(l)}_{Q}\cos\varphi^{(l)}_{P}) +\alpha
E_{{\rm J},l}[1-\cos(2\varphi^{(l)}_{P}+2\pi f_{l})]
$
with the redefined phases
$\varphi^{(l)}_{P}=(\varphi^{(l)}_{1}+\varphi^{(l)}_{2})/2$,
$\varphi^{(l)}_{Q}=(\varphi^{(l)}_{1}-\varphi^{(l)}_{2})/2$, and
reduced bias magnetic flux $f_{l}=\Phi_{\rm e}^{(l)}/\Phi_{0}$.
The effective masses are $M_{Q,l}=2(\Phi_{0}/2\pi)^2C_{{\rm J},l}$
and $M_{P,l}=(1+2\alpha)M_{Q,l}$, which correspond to the
effective momenta $P_{Q,l}=-i\hbar\partial/\partial
\varphi^{(l)}_{Q}$ and $P_{P,l}=-i\hbar\partial/\partial
\varphi^{(l)}_{P}$. The capacitances in the $l$th qubit loop
satisfy the condition $C^{(l)}_{1}=C^{(l)}_{2}=C_{{\rm J},l}$ and
$C^{(l)}_{3}=\alpha C_{{\rm J},l}$. The Hamiltonian
$H_{\rm D}^{(l)}= -(A_{l}/2)\left(I^{(l)}_{3}+i\beta
P_{P,l}\right)e^{-i\omega^{(l)}_{c}t}+{\rm H.c.}$ 
represents the interaction between the $l$th qubit and its TDMF.
Here, the parameter $\beta=
2\pi\alpha\omega^{(l)}_{c}/[\Phi_{0}(1+2\alpha)]$,  and
$I^{(l)}_{3}=-(2\pi\alpha E_{{\rm
J},l}/\Phi_{0})\sin(2\varphi^{(l)}_{P}+2\pi f_{l})$ is the
supercurrent through the smaller junction of the $l$th qubit
without applying the TDMF. So a TDMF-controlled single-qubit
rotation can be realized by the Hamiltonian $H_{\rm D}^{(l)}$. The
qubit-qubit interaction $H_{lm}$, controlled by one of the TDMFs
($\Phi^{(1)}_{\rm e}(t)$ {\it or} $\Phi^{(2)}_{\rm e}(t)$), can be
described by
$H_{lm}=-\beta_{l}I^{(m)}e^{-i\omega^{(l)}_{c}t}\cos(2\varphi^{(l)}_{P}+2\pi
f_{l})+{\rm H.c.}$,
where $\beta_{l}=M(2\pi/\Phi_{0})^2 (A_{l}C_{l}E_{{\rm J},
l}/2C_{{\rm J}, l})$, and
$I^{(m)}=C_{m}\sum_{i}(I^{(m)}_{0i}/C^{(m)}_{i})\sin\varphi^{(m)}_{i}$
is the qubit loop-current of the $m$th qubit without an applied
TDMF, and $C^{-1}_{m}=\sum_{i=1}^{3}(C^{(m)}_{i})^{-1}$. However,
the qubit-qubit interaction $H_{C}$, controlled by simultaneously
applying two TDMFs ($\Phi^{(1)}_{\rm e}(t)$  and $\Phi^{(2)}_{\rm
e}(t)$) through the two qubits, respectively, is
$H_{C}=B\prod_{l=1}^{2}
\Phi^{(l)}_{e}(t)\cos(2\varphi^{(l)}_{p}+2\pi f_{l})$
with $B=M\left(2\pi/\Phi_{0}\right)^4(C_{1}C_{2}E_{{\rm J},
1}E_{{\rm J}, 2}/C_{{\rm J},1}C_{{\rm J},2})$.
The Hamiltonian $H_{A}=MI^{(1)}I^{(2)}$ denotes an always-on
interaction between the two flux qubits, without applying the
TDMF.

In the two-qubit basis
$\{|e_{1}\rangle,\,|g_{1}\rangle\}\otimes\{|e_{2}\rangle,\,|g_{2}\rangle\}$,
where $|g_{l}\rangle$ and $|e_{l}\rangle$ are the two lowest
eigenstates (ground and first excited states) of $H_{l}$,
Eq.~(\ref{eq:1}) can become~\cite{neglect1}
\begin{eqnarray}\label{eq:2}
H&=&\sum_{l=1}^{2}\frac{1}{2}\hbar\,\omega_{l}\sigma^{(l)}_{z}-\sum_{l=1}^{2}
\left(\kappa_{l}\sigma^{(l)}_{+}e^{-i\omega^{(l)}_{c}t}+{\rm
H.c.}\right)\nonumber\\
&-& \sum_{l\neq
m=1}^{2}\left(\Omega^{(1)}_{lm}\,\sigma^{(l)}_{+}\sigma^{(m)}_{-}+{\rm
H.c.}\right)
\left(e^{i\omega^{(l)}_{c}t}+e^{-i\omega^{(l)}_{c}t}\right)\nonumber \\
&-&  \sum_{l\neq
m=1}^{2}\left(\Omega^{(2)}_{lm}\,\sigma^{(l)}_{+}\sigma^{(m)}_{+}e^{-i\omega^{(l)}_{c}t}+{\rm
H.c.}\right)\nonumber\\
&+&\left(\lambda_{1}\,\sigma^{(1)}_{+}\sigma^{(2)}_{-}+\lambda_{2}\,\sigma^{(1)}_{+}\sigma^{(2)}_{+}
+{\rm H.c.}\right)\, .
\end{eqnarray}
Here, the terms
$\kappa^{*}_{l}\,\sigma^{(l)}_{+}e^{i\omega^{(l)}_{c}t}$ and
$\Omega^{(2)}_{lm}\,\sigma^{(l)}_{+}\sigma^{(m)}_{+}e^{i\omega^{(l)}_{c}t}$,
as well as their complex conjugates, have been neglected by
considering energy conservation. The operators of the $l$th qubit
are defined as $\sigma_{z}^{(l)}=|e_{l}\rangle\langle
e_{l}|-|g_{l}\rangle\langle g_{l}|$,
$\sigma_{+}^{(l)}=|e_{l}\rangle\langle g_{l}|$, and
$\sigma_{-}^{(l)}=|g_{l}\rangle\langle e_{l}|$. The qubit
frequency $\omega_{l}$ in Eq.~(\ref{eq:2}) can be expressed as
$\omega_{l}=\sqrt{2I^{(l)}(\Phi^{(l)}_{\rm
e}-\Phi_{0}/2)^2+t_{l}^2}$ with the loop-current $I^{(l)}$ and the
bias flux $\Phi^{(l)}_{\rm e}$. Here, the parameter $t_{l}$
denotes the tunnel coupling between two wells in the $l$th
qubit~\cite{orlando}. The controllable coupling constants are
$\kappa_{l}=A_{l}\langle e_{l}|(I^{(l)}_{3}+i\beta
P_{P,l})|g_{l}\rangle/2$, $\Omega^{(1)}_{lm}=A_{l}\beta_{l}\langle
e_{l},g_{m}|I^{(m)}\cos(2\varphi^{(l)}_{P}+2\pi
f_{l})|g_{l},e_{m}\rangle/2$, and
$\Omega^{(2)}_{lm}=A_{l}\beta_{l}\langle
e_{l},e_{m}|I^{(m)}\cos(2\varphi^{(l)}_{P}+2\pi
f_{l})|g_{l},g_{m}\rangle/2$. The hard-to-control parameters are
$\lambda_{1}=M\langle e_{1}, g_{2}|I^{(1)}I^{(2)}|g_{1},
e_{2}\rangle$, and $\lambda_{2}=M\langle e_{1},
e_{2}|I^{(1)}I^{(2)}|g_{1}, g_{2}\rangle$. It is not difficult to
derive that $\kappa_{l}=\Omega^{(1)}_{lm}=\Omega^{(2)}_{lm}=0$
when no TDMFs. Then, since both bias magnetic fluxes $f_{l}$ are
near $1/2$ (the optimal point is at $f_{l}=1/2$), the
Hamiltonian~(\ref{eq:2}) can revert to the case in
Refs.~\cite{izmalkov,majer},  where the Pauli operators are
defined by the states of the two potential wells.

{\it Decoupling mechanism and logic gates.---} We assume that the
two qubits work at the fixed frequencies $\omega_{1}$ and
$\omega_{2}$, which mean that their reduced bias magnetic fluxes
$f_{l}\,(l=1,\,2)$ and the frequency difference
$\Delta=\omega_{1}-\omega_{2}$ remain fixed. If the detuning
$\Delta$ is initially chosen to be sufficiently large (such that
it satisfies the condition: $|\Delta|\gg
|\lambda_{1}|/\hbar=|\lambda_{2}|/\hbar=|\lambda|/\hbar$), then
the two qubits can be approximately treated as two decoupled
subsystems ~\cite{mcdermott} when the TDMFs are not applied.

By applying the TDMF with the frequency-matching condition
$\omega^{(l)}_{c}=\omega_{l}$, we can easily derive from
Eq.~(\ref{eq:2}) that any single-qubit operation of the $l$th
qubit can be performed via the dynamical evolution
$U^{(l)}_{c}(\theta_{l},\phi_{l})=
\exp[i\theta_{l}(e^{-i\phi_{l}}\sigma^{(l)}_{-} +{\rm H.c.})]$.
Here, $\theta_{l}=|\kappa_{l}|\tau/\hbar$ depends on the Rabi
frequency $|\kappa_{l}|/\hbar$ and duration $\tau$; $\phi_{l}$ is
related to the TDMF phase applied to the $l$th qubit. For example,
$\pi/2$ rotations of the $l$th qubit around the $x$ or $y$ axes
can be performed by $U^{(l)}_{c}(\theta_{l},\phi_{l})$, with
$\tau=\hbar\pi/4|\kappa_{l}|$ and $\phi_{l}=\pi$ or $\pi/2$.
Unless specified otherwise, hereafter, we work in the interaction
picture, and all non-resonant terms have been neglected because
their contributions to the transitions between different states
are negligibly small~\cite{louisell}.

To couple two qubits with the assistance of the TDMF: i) a TDMF
needs to be applied through one of the qubits, and its frequency
should be equal to the detuning (or summation) of the two qubit
frequencies; ii) the reduced bias flux~\cite{symmetry} on the
qubit, which is addressed by the TDMF, should be near but not
equal to $1/2$. Without loss of generality, below, the TDMF is
assumed to be always applied through the first qubit, so the bias
of the first qubit is $f_{1}=1/2\pm \epsilon$, with small
$\epsilon$; however, the bias for the second qubit is taken as
$f_{2}=1/2$.

Considering two new frequency-matching conditions in
Eq.~(\ref{eq:2}), produces two different kinds of Hamiltonians for
implementing two-qubit operations with the assistance of TDMF. One
is $H_{1}=\Omega^{(1)}_{12}\sigma^{(1)}_{+}\sigma^{(2)}_{-}+{\rm
H.c.}$, with the condition:
$\omega_{1}-\omega_{2}\pm\omega^{(1)}_{c}=0$. Here, the sign is
positive (negative) when $\Delta < 0$ ($\Delta > 0$). Another one
is $H_{2}= \Omega^{(2)}_{12}\sigma^{(1)}_{+}\sigma^{(2)}_{+}+{\rm
H.c.}$, when the frequencies satisfy the condition:
$\omega_{1}+\omega_{2}-\omega^{(1)}_{c}=0$. Using the Hamiltonian
$H_{1}$ and $H_{2}$, two qubit gates can be implemented. For
example, a TDMF is applied through the first qubit with its
frequency $\omega^{(1)}_{c}$, satisfying the condition
$\omega_{1}-\omega_{2}-\omega^{(1)}_{c}=0$ (hereafter, we assume
$\omega_{1}>\omega_{2}$ without loss of generality). The
Hamiltonians $H_{1}$ can be transformed to
$H_{1}=H_{XY}=-(|\Omega^{(1)}_{12}|/2)(\sigma^{(1)}_{x}\sigma^{(2)}_{x}
+\sigma^{(1)}_{y}\sigma^{(2)}_{y})$, when the phase of
$\Omega^{(1)}_{12}$ is set to $\pi$ by an applied TDMF. Then an
ISWAP gate~\cite{schuch}, denoted by $U_{\rm iS}$, can be realized
by $H_{XY}$ with an evolution time
$t=\pi\hbar/(2|\Omega^{(1)}_{12}|)$. The CNOT gate can be
constructed by combining the ISWAP gate with a few single-qubit
operations.

Experimentally, it is found that the always-on coupling strength
$|\lambda|/h$ is about several hundred MHz, e.g. $|\lambda|/h\sim
0.4$ GHz in Refs.~\cite{izmalkov,majer}, the detuning
$\Delta/2\pi$ can be $\sim 1$ to $10$ GHz. This means that the
ratio $|\lambda|/\hbar\Delta$ cannot be infinitesimally small, and
the always-on interaction needs to be considered when all TDMFs
are switched off.  Up to the first order in
$|\lambda|/\hbar\Delta$, the effect of the always-on interaction,
without the TDMF, can be described by an effective Hamiltonian $
H_{E}=(|\lambda|^2/\hbar\Delta)[|e_{1}\rangle\langle
e_{1}|\otimes|g_{2}\rangle\langle g_{2}|-|g_{1}\rangle\langle
g_{1}|\otimes |e_{2}\rangle\langle e_{2}|]$.  Here, fast
oscillating terms have been neglected.

{\it Entangled states and tomographic measurement.---} Entangled
states can be easily generated in this circuit. For example, if
the first and second qubits are initially in the excited state
$|e_{1}\rangle$ and the ground state $|g_{2}\rangle$, then using
the Hamiltonian $H_{1}$, the system evolves to
$|\Psi_{1}(t)\rangle=\cos(|\Omega^{(1)}_{12}|t/\hbar)|e_{1},g_{2}\rangle-\!\!
ie^{-i\theta}\sin(|\Omega^{(1)}_{12}|t/\hbar)|g_{1},e_{2}\rangle$
with the phase $\theta$ of $\Omega^{(1)}_{12}$. It is obvious that
the Bell states $|\psi_{\pm}\rangle=(|g_{1},e_{2}\rangle \pm
|e_{1},g_{2}\rangle)/\sqrt{2}$ can be generated with
$t_{1}=\hbar\pi/(4|\Omega^{(1)}_{12}|)$ by setting the TDMF such
that $\theta=\pi/2$ or $3\pi/2$. Similarly, if both qubits are in
the ground states $|g_{1}\rangle$ and $|g_{2}\rangle$, then
another two Bell states $|\Psi_{\pm}\rangle=(|g_{1},g_{2}\rangle
\pm |e_{1},e_{2}\rangle)/\sqrt{2}$ can also be obtained with
$t_{2}=\hbar\pi/(4|\Omega^{(2)}_{12}|)$ by setting the phase
$\theta^{\prime}$ of $\Omega^{(2)}_{12}$ as $\pi/2$ or $3\pi/2$
through the Hamiltonian $H_{2}$.

State tomography allows us to experimentally determine a quantum
state~\cite{liu}. Qubit state tomography can be implemented by
measuring the supercurrent through the qubit loop, which is
inductively coupled to, e.g., a dc SQUID magnetometer or
high-quality tank circuit~\cite{izmalkov}. For the $l$th qubit in
the qubit basis $\{|g_{l}\rangle,\,|e_{l}\rangle\}$, its
loop-current operator can be written~\cite{jq} as
\begin{eqnarray}\label{eq:6}
\hat{I}^{(l)}&\equiv
&\hat{I}^{(l)}_{x}=a_{l}\,\sigma^{(l)}_{z}+b_{l}\,|e_{l}\rangle\langle
g_{l}|+b^{*}_{l} \,|g_{l}\rangle\langle e_{l}|
\end{eqnarray}
with $a_{l}=\langle e_{l}|I^{(l)}|e_{l}\rangle$ and $b_{l}=\langle
e_{l}|I^{(l)}|g_{l}\rangle$, when the bias $f_{l}$ is near (but
not equal to)  $1/2$. However, at the optimal point $f_{l}=1/2$,
the supercurrent operator in Eq.~(\ref{eq:6}) can be
reduced~\cite{jq} to
$\hat{I}^{(l)}\equiv\hat{I}^{(l)}_{x}=b_{l}\,\sigma^{(l)}_{x}$,
with a real number $b_{l}$ and the Pauli operator
$\sigma^{(l)}_{x}=|e_{l}\rangle\langle g_{l}|+|g_{l}\rangle\langle
e_{l}|$.

If the simultaneous joint measurement of two qubits  can be
performed in flux qubit circuits as in phase
circuits~\cite{mcdermott}, then single qubit operations are enough
to realize the fifteen different measurements~\cite{liu} on the
two-qubit states $\rho=(1/4)\sum_{i
,j}r_{i,j}\,\sigma^{(1)}_{i}\otimes \sigma^{(2)}_{j}$, with the
Pauli operators $\sigma_{i}^{(l)}$ ($i,\,j=x,\,y,\,z$ and
$l=1,\,2$) and the identity operator $\sigma^{(l)}_{0}$, where
$r_{00}=1$ by normalization. The loop-current operator for the
first qubit is given in Eq.~(\ref{eq:6}) by setting $l=1$ due to
the assumption $f_{1}\neq 1/2$, but it is reduced to
$I^{(2)}_{x}=b_{2}\sigma^{(2)}_{x}$ for the second qubit with
$f_{2}=1/2$. So the fifteen measurements on state $\rho$ are given
as $I^{(1)}_{i}$ and $b_{2}\sigma^{(2)}_{j}$ (denoted as
single-qubit measurements),  as well as $b_{2} I^{(1)}_{i}\otimes
\sigma^{(2)}_{j}$ (called two-qubit or joint measurements), with
$i,\,j=x,\,y,\,z$, $I^{(1)}_{y}=Y^{\dagger}_{1}I^{(1)}_{x}Y_{1}$,
and $I^{(1)}_{z}=Z^{\dagger}_{1}I^{(1)}_{x}Z_{1}$. It is clear
that three measurements ($I^{(1)}_{x}$, $b_{2}\sigma^{(2)}_{x}$,
and $b_{2}I^{(1)}_{x}\otimes\sigma^{(2)}_{x}$) on the input
two-qubit state $\rho$ can be directly performed. Other twelve
measurements can be equivalently obtained by measuring
($I^{(1)}_{x}$, $b_{2}\sigma^{(2)}_{x}$, or both of them at the
same time) on the rotated state $\rho$. For example,  $\pi/2$
single-qubit rotations $Y_{1}$ around the $y$ axis for the first
qubit and $Z_{2}$ around the $z$ axis for the second qubit are
simultaneously performed on the state $\rho$, then the measurement
$b_{2} I^{(1)}_{x}\otimes \sigma^{(2)}_{x}$ on the rotated state
$Y_{1}Z_{2}\rho Z^{\dagger}_{2}Y^{\dagger}_{1}$ is equivalent to
the measurement $b_{2} I^{(1)}_{y}\otimes \sigma^{(2)}_{y}$ on the
original state $\rho$. Similarly, other joint measurements can
also be obtained. Finally, for the fifteen measured results, we
solve a set of equations for the parameters $r_{ij}$, and a
two-qubit state is reconstructed.

If only a single-qubit measurement can be made at a time, besides
six single-qubit measurements mentioned above, a suitable nonlocal
two-qubit operation~\cite{liu} is required to obtain the
coefficients (e.g., $r_{y,z}$) of the nine joint measurements on
the state $\rho$. Here, this is an ISWAP gate $U_{iS}$, which can
be implemented as described above. For example, if an operation
$U_{\rm iS}$ is made on the input state $\rho$, then the
loop-current of the second qubit should be $\langle
I^{(2)}_{x}\rangle={\rm Tr}(U_{\rm iS}\,\rho\,U^{\dagger}_{\rm
iS}\,I^{(2)}_{x})=-b_{2}{\rm
Tr}(\rho\,\sigma_{y}^{(1)}\otimes\sigma_{z}^{(2)})=-b_{2}r_{y,z}$,
and the coefficient $r_{y,z}$ is determined. Combining the ISWAP
gate and single-qubit operations for two qubits, the coefficients
of other eight joint measurements can also be determined by only
measuring the loop-current $I^{(2)}_{x}$.

\begin{figure}[tbp]
\includegraphics[width=28mm]{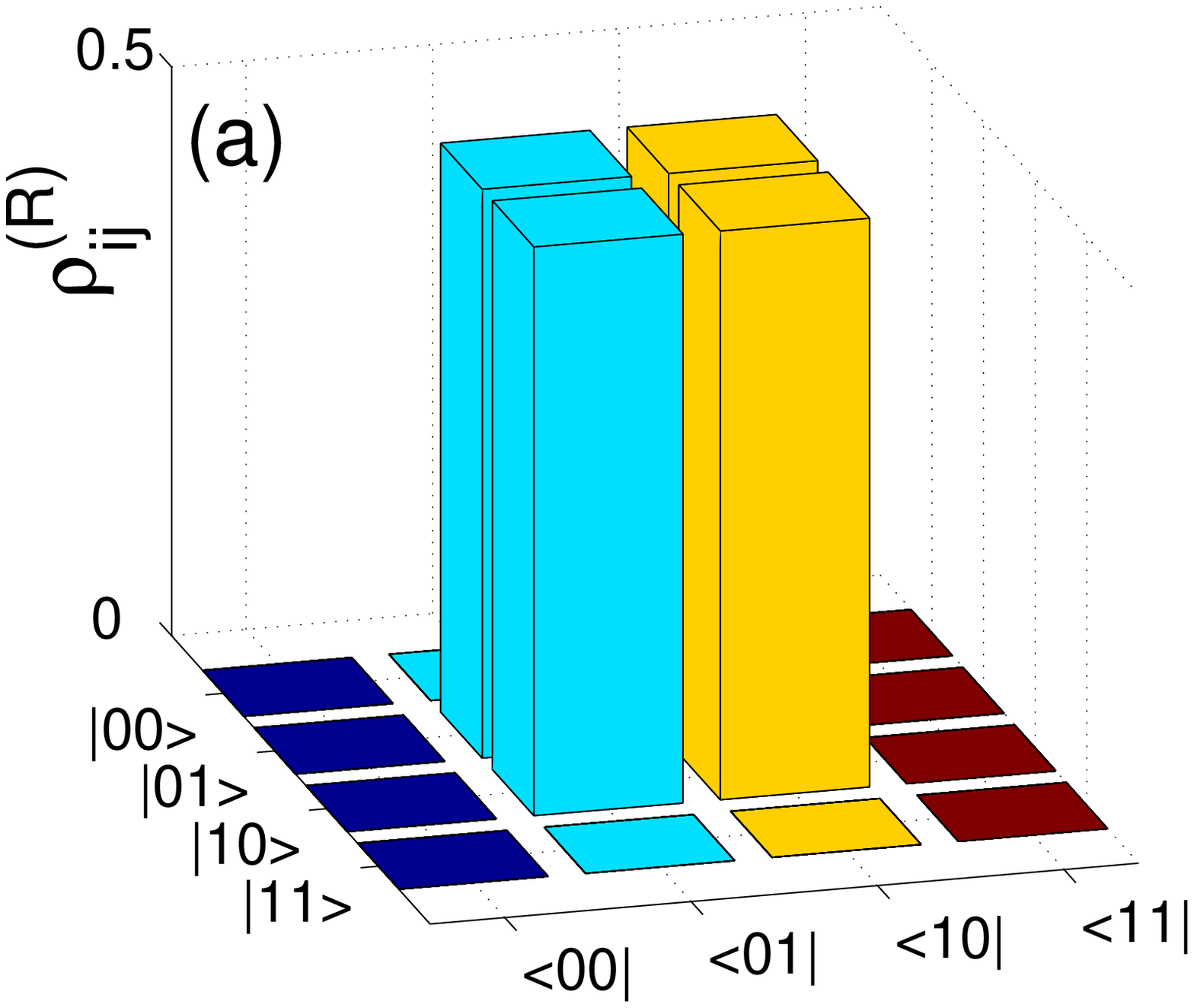}
\includegraphics[width=28mm]{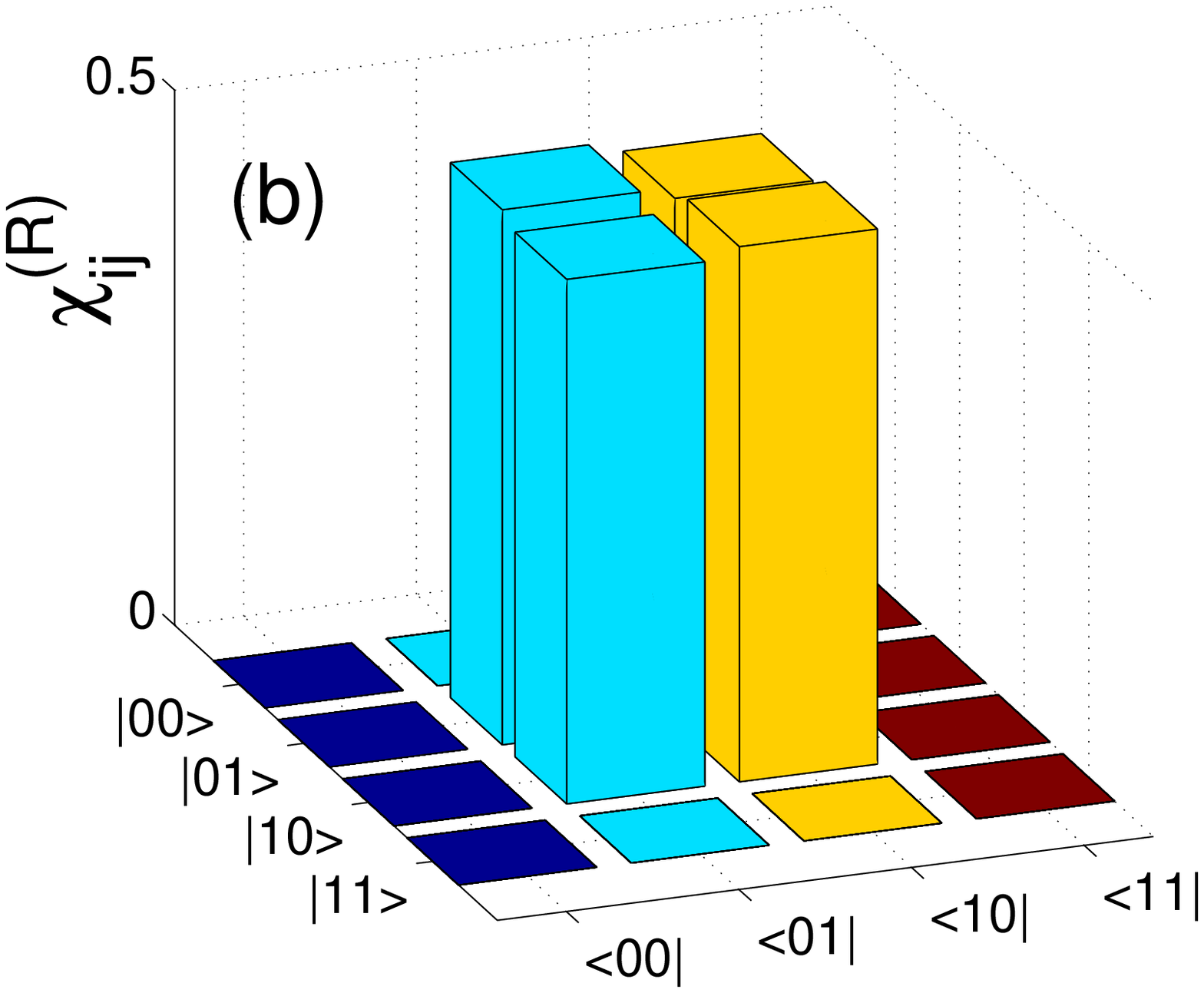}
\includegraphics[width=28mm]{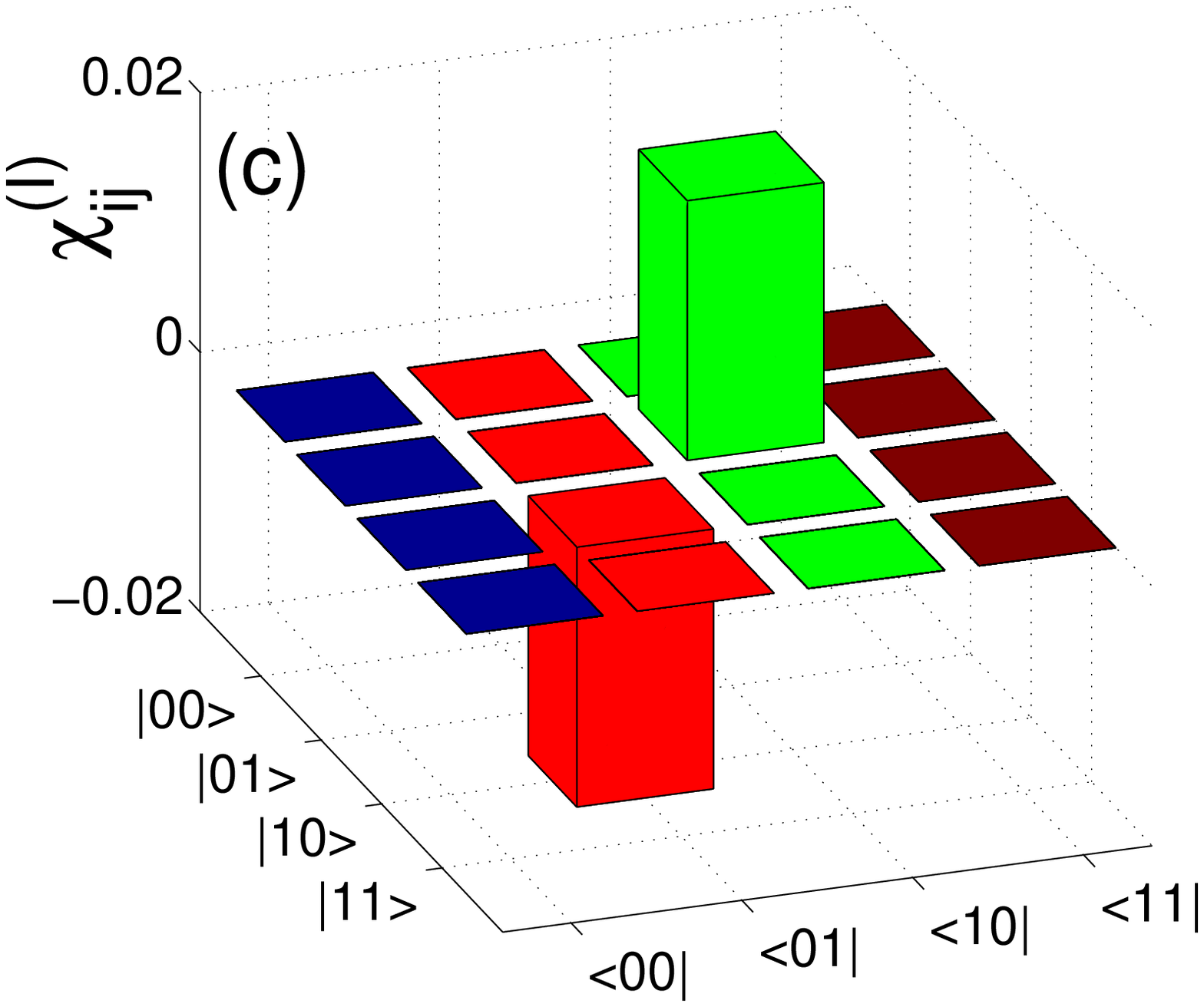}
\caption[]{(Color online) A schematic representation of
tomographically measured state $\rho$ [with only real part in (a)]
for a completely decoupled system, as well as a state $\chi$ [with
real part in (b) and imaginary part in (c)] for a two-qubit system
with large detuning---after considering a duration $t$ before
measuring the created state $|\psi_{+}\rangle$.}\label{fig2}
\end{figure}

Tomographically measured states are different for completely
decoupled (CD) and large detuning (LD) two-qubit systems after
two-qubit states are created, if we consider a duration $t$ before
measuring the generated two-qubit states. As an example, a
schematic representation of a Bell state $|\psi_{+}\rangle$ is
given in Fig.~\ref{fig2} for the above two cases. There is only
the real part for the reconstructed state
$\rho=|\psi_{+}\rangle\langle \psi_{+}|$ in the CD system, shown
in Fig.~\ref{fig2}(a). However, due to the effect of $H_{E}$ for
the LD system, the reconstructed state
$\chi=e^{-iH_{E}t/\hbar}|\psi_{+}\rangle\langle
\psi_{+}|e^{iH_{E}t/\hbar}$ includes both real and imaginary
parts, shown in Fig.~\ref{fig2}(b) and ~\ref{fig2}(c),
respectively. In Fig.~\ref{fig2}, we consider a longer duration
$t\sim 10^{-9}$ s; the detuning and the coupling constant are,
e.g., $\Delta\sim 5$ GHz, and $|\lambda|/h\sim 0.4$ GHz. So if we
consider the always-on interaction effect, the relative error with
these parameters is $\sim 0.08 \,$ for the non-diagonal parts of
the reconstructed CD state $|\psi_{+}\rangle$. Here, the qubit
free-evolution is neglected. In practice, considering unavoidable
environmental effects and statistical errors, the experimentally
measured data should be further optimized by other
methods~\cite{tomography} .

{\it Conclusions.---} The controllable coupling of two
inductively-coupled flux qubits can be realized, when the large
detuning condition is satisfied, by the frequency of the TDMF
matching/mismatching to the detuning (or summation)  of the two
qubit frequencies; not by changing qubit biases (e.g., as in
Ref.~\cite{mcdermott}). Our proposal is also different from the
coupling/decoupling method by using dressed states~\cite{rigetti}.
We emphasize that the deviation $\epsilon$ from the optimal point
$1/2$ for the reduced bias $f_{1}$ through the first qubit will
make the decoherence time $T_{2}$ short. However, our proposal can
work for a small deviation, e.g., $\epsilon\sim 10^{-4}$, in which
$T_{2}\gtrsim 20$ ns in Refs.~\cite{bertet} or $T_{2}\gtrsim 100$
ns with spin-echo signals~\cite{bertet}. At this point, the qubit
coupling constant $\Omega^{(i)}_{12}\,(i=1,\,2)$  can
reach~\cite{symmetry} about several hundred MHz. Based on
numerical estimates~\cite{jq}, the longest time for the
single-qubit operation $Z_{l}$ is less than $5$ ns, so the
tomographic measurements can be performed within $T_{2}$. Our
proposal can be scalable to a chain of many inductively coupled
flux qubits, if all of qubits satisfy the large detuning
condition. We need to note: i) we can use one LC circuit as a
common information bus to couple many qubits, with the qubit-bus
coupling controlled by externally variable frequencies~\cite{bus};
ii) this circuit can be modified to work at the optimal point;
iii) this method using frequency-controlled couplings can be
applied to control one-junction flux qubits. It can also be
modified to control phase, charge-flux, and charge qubits.

We thank J.Q. You, Y. Nakamura, Y.A. Pashkin, O. Astafiev, and  K.
Harrabi for helpful discussions. This work was supported in part
by the NSA and ARDA under AFOSR contract No. F49620-02-1-0334, and
by the NSF grant No. EIA-0130383.


\vspace{-0.5cm}


\begin{thebibliography}{99}

\vspace{-0.5cm}










\bibitem{physics} J. Q. You and F. Nori, Phys. Today 58 No.11, 42(2005)

\bibitem{pashkin}Y.A. Pashkin {\it et al.}, Nature {\bf 421}, 823 (2003); T.
Yamamoto {\it et al.}, {\it ibid.} {\bf 425}, 941 (2003).


\bibitem{izmalkov}A. Izmalkov {\it et al.}, Phys. Rev. Lett. {\bf
93}, 037003 (2004); M. Grajcar {\it et al.}, Phys. Rev. B {\bf
72}, 020503(R) (2005).

\bibitem{xu}H. Xu {\it et al.}, Phys. Rev. Lett. {\bf 94}, 027003
(2005).

\bibitem{berkley} A.J. Berkley {\it et al.}, Science {\bf 300}, 1548
(2003).

\bibitem{majer}J.B. Majer {\it et al.}, Phys. Rev. Lett. {\bf 94},
090501 (2005).

\bibitem{mcdermott}R. McDermott {\it et al.}, Science {\bf 307},
1299 (2005).


\bibitem{blais}A. Blais {\it et al.}, Phys. Rev. Lett. {\bf 90}, 127901
 (2003); A. Blais {\it et al.}, Phys. Rev. A {\bf 69}, 062320 (2004).

\bibitem{cleland}A.N. Cleland and M.R. Geller, Phys. Rev. Lett. {\bf
93}, 070501 (2004).

\bibitem{makhlin}Y. Makhlin {\it et al.}, Rev. Mod. Phys. {\bf
73}, 357 (2001).

\bibitem{you}J.Q. You {\it et al.}, Phys. Rev. Lett. {\bf 89},
197902 (2002); Y.X. Liu {\it et al.}, Europhys. Lett. {\bf 67},
941 (2004). L.F. Wei {\it et al.}, {\it ibid.} {\bf 67}, 1004
(2004); Phys. Rev. B {\bf 71}, 134506 (2005).


\bibitem{averin}D.V. Averin and C. Bruder, Phys. Rev. Lett. {\bf
91}, 057003 (2003).

\bibitem{plourde}B.L.T. Plourde {\it et al.}, Phys. Rev. B {\bf 70},
140501(R) (2004).



\bibitem{neglect}Here, we have neglected the self-interaction term of the external
magnetic fields $H_{\rm self}=\sum_{l=1}^{2}(\alpha
C^{(l)}_{J}/2)(\partial \Phi^{(l)}_{e}(t)/\partial t)^2$ and the
self-inductance energy terms for two qubits.


\bibitem{neglect1}In the derivation of Eq.~(\ref{eq:2}),
we have discarded the two-fields-controlled qubit-qubit
interaction term
$H_{C}=B(\chi_{1}\sigma^{(1)}_{+}+\chi^{*}_{1}\sigma^{(1)}_{-})
(\chi_{2}\sigma^{(2)}_{+}+\chi^{*}_{2}\sigma^{(2)}_{-})
\Phi^{(1)}_{e}(t)\Phi^{(2)}_{e}(t)$ with $\chi_{l}=\langle
1|\cos(2\varphi^{(l)}_{p}+2\pi f_{l})|0\rangle\,\, (l=1,\,2)$. In
principle, the qubit-qubit interaction can also be controlled by
two external fields, one for each qubit loop.

\bibitem{orlando}J.E. Mooij {\it et al.}, Science {\bf 285}, 1036
(1999); T.P. Orlando {\it et al.}, Phys. Rev. B {\bf 60}, 15398
(1999).

\bibitem{louisell}W.H. Louisell, {\it Quantum Statistical Properties
of Radiation} (John Wiley \& Sons, New York, 1973).

\bibitem{symmetry}Y.X. Liu {\it et al.}, Phys. Rev. Lett. {\bf 95},
087001 (2005).

\bibitem{schuch}N. Schuch and J. Siewert, Phys. Rev. A {\bf 67},
032301 (2003).


\bibitem{liu}Y.X. Liu {\it et al.},  Europhys. Lett. {\bf
67}, 874 (2004); Phys. Rev. B {\bf 72}, 014547 (2005).


\bibitem{jq}J.Q. You {\it et al.},  Phys. Rev. B {\bf
71}, 024532 (2005).






\bibitem{tomography}M. Paris and J. Rehacek, {\it Quantum State
Estimation}, Lecture Notes in Physics, {\bf 649} (Springer,
Berlin, 2004).

\bibitem{rigetti}C. Rigetti {\it et al.}, Phys. Rev. Lett. {\bf
94}, 240502 (2005).

\bibitem{bertet}P. Bertet {\it et al.}, Phys. Rev. Lett. {\bf
95}, 257002 (2005).


\bibitem{bus}Y. X. Liu {\it et al.}, cond-mat/0509236.
\end{thebibliography}
\end{document}